\documentclass{PoS}

\usepackage{multirow}
\usepackage{amsmath}
\DeclareMathAlphabet{\mathcal}{OMS}{cmsy}{m}{n}
\usepackage{lineno}
\title{Probing Galactic Diffuse TeV Gamma-Ray Emission with the HAWC Observatory}

\ShortTitle{Galactic Diffuse Emission with HAWC}

\author{\speaker{Hao Zhou}\\
        Los Alamos National Laboratory\\
        E-mail: \email{hao@lanl.gov}}

\author{Chang Dong Rho\\
        University of Rochester\\
        E-mail: \email{crho2@ur.rochester.edu}}

\author{Giacomo Vianello\\
        Stanford University\\
        E-mail: \email{giacomov@stanford.edu}}
        
\author{for the HAWC Collaboration\thanks{For a complete author list, see http://www.hawc-observatory.org/collaboration/icrc2017.php}}

\abstract{

Galactic diffuse TeV gamma-ray emission is produced by the interaction of high-energy cosmic-ray particles with matter and radiation in our Galaxy. The measurement of Galactic diffuse TeV gamma-ray emission would provide strong constraints on the acceleration and propagation of Galactic cosmic rays. The High Altitude Water Cherenkov (HAWC) Observatory, located in central Mexico at 4100 m above sea level, is sensitive to gamma rays between a few hundreds GeV and $\sim100$ TeV. Thanks to its large field of view of 2 steradians and excellent background estimation, HAWC has a unique capability to observe large extended sources such as Galactic diffuse emission. We present the analysis strategy to detect the Galactic diffuse emission with HAWC, including techniques to remove contaminations from localized sources.

}

\FullConference{35th International Cosmic Ray Conference --- ICRC2017\\
		10--20 July, 2017\\
		Bexco, Busan, Korea}

\begin{document}

\section{Introduction}

Over one hundred years after their discovery, the origin of Galactic cosmic rays remains unclear. Supernova remnants (SNRs) and pulsars are postulated as the sources of Galactic cosmic rays. The Galactic diffuse gamma-ray emission is produced by the interaction of high-energy cosmic-ray particles with matter and radiation in our Galaxy. Studying the Galactic diffuse emission, we learn about propagation of cosmic rays and can infer what are the most likely sources of cosmic-ray acceleration. At TeV energies, there are two major mechanisms that produce Galactic diffuse emission: $\pi^0$ decays and inverse Compton scattering, referred to as hadronic and leptonic processes. In the hadronic process, gamma rays are produced by secondary $\pi^0$ decays from cosmic-ray hadrons interacting with the interstellar medium, while in the leptonic process, gamma rays are from high-energy electrons inverse Compton scattering off the interstellar radiation fields. The Galactic diffuse gamma-ray emission provides excellent constraints on the acceleration and propagation of high-energy cosmic rays within our Galaxy.

The Galactic diffuse TeV gamma-ray emission has been studied by the Milagro experiment \cite{milagro_GDE}, ARGO-YBJ \cite{argo_GDE} and H.E.S.S. \cite{hess_GDE}. At this energy regime, the gamma-ray flux from resolved sources dominates the total observed flux. Therefore, unresolved sources potentially contribute a large fraction of the total observed flux as well. Milagro and ARGO-YBJ, as all-sky survey-type instruments, have high duty cycle and large field of view, which are favorable for the observations of large structures. However, they have higher source detection thresholds thus a larger contribution from unresolved sources is expected. On the other hand, H.E.S.S., utilizing the imaging atmospheric Cherenkov technique, has excellent angular resolution and gamma-hadron separation quality. But the small field of view limits its ability on detecting large structures such as the inverse Compton emission component of the Galactic diffuse emission. HAWC is a second-generation water Cherenkov experiment, which has an order of magnitude better sensitivity than its predecessor Milagro. With better angular resolution and gamma-hadron separation than its predecessors, as well as high duty cycle and large field of view, HAWC has a unique ability to study the Galactic diffuse TeV gamma-ray emission.

In these proceedings, we will demonstrate the method of analysis on Galactic diffuse emission with HAWC data and discuss possible contributions from unresolved sources under the detection threshold of HAWC. Results of this analysis will be presented at ICRC2017.

\section{HAWC Observations}

The High Altitude Water Cherenkov (HAWC) gamma-ray observatory, located in central Mexico at 4100 m above sea level, is sensitive to gamma rays between a few hundreds GeV and $\sim100$ TeV. The detector consists of 1200 photomultiplier tubes (PMTs) installed in 300 water Cherenkov detectors. HAWC has been fully operational since March 2015. %Figure \ref{fig:sigma} shows the significance map from 17 months of data with the complete HAWC detector. 
The 2HWC catalog has been published based on 17 months of data collected between November 2015 and June 2016 \cite{2hwc}. A total of 39 sources are identified in this all-sky survey, with 34 of them within $10^\circ$ from the Galactic plane. In this work we use the same dataset to search for underlaying diffuse emission after 2HWC sources subtracted and to estimate the contribution from unidentified sources.

Gamma-ray events detected by HAWC are classified by size in nine analysis bins $\mathcal{B}$, depending on the fraction of active PMTs triggered by a gamma-ray event. This fraction is a proxy for the energy of the gamma-ray photon and the nine analysis bins $\mathcal{B}$ are proxy for the energy bins. 

%\begin{figure}[htpb]
%\begin{center}
%\includegraphics[width=0.9\linewidth]{sigmap.pdf}
%\caption{Significance map with 17 months of data of HAWC.}
%\label{fig:sigma}
%\end{center}
%\end{figure}

\section{Analysis Methods}

One important question that we try to answer with this study is that how the diffusion emission distributes as a function of Galactic latitude or longitude. In order to investigate the longitudinal and latitudinal profiles, we first generate an all-sky map of flux density (flux per steradian) by comparing data and simulations using a HEALPix pixelization scheme \cite{healpix}, with a mean pixel size of less than $0.12^\circ$ ($N_{side} = 512$). This map is constructed following these steps;

\begin{enumerate}
\item Data: construct an all-sky excess map $E_\mathcal{B}$ in each analysis bin $\mathcal{B}$ with 2HWC sources removed by
\begin{enumerate}
  \item subtracting the background map using the direct integration background estimation method \cite{di} from the raw data counts map;
   \item subtracting the contribution from 2HWC sources in each pixel using the locations and spectra reported in \cite{2hwc} convolved with the detector response.
\end{enumerate}
\item Simulation: construct an all-sky expected counts map $M_\mathcal{B}$ in each analysis bin $\mathcal{B}$, as the detector response is a function of source declination,  for a simulated source with a power-law spectrum assumption of,
  \begin{enumerate}
    \item a flux normalization of $1.64\times 10^{-13} (\textrm{TeV}\,\textrm{cm}^2\,\textrm{s})^{-1}$ at 7\;TeV (Crab flux),
    \item a spectral index of 2.75, which was chosen to match the cosmic-ray spectrum around 10 TeV \cite{milagro_GDE}.
  \end{enumerate}
  
\item Calculate the flux density $S$ at 7 TeV in each pixel by comparing data to simulations, 

\begin{equation}
S = \frac{\sum_{\mathcal{B}=1}^9 (E_\mathcal{B} \times GHW_\mathcal{B})}{\sum_{\mathcal{B}=1}^9 (M_\mathcal{B} \times GHW_\mathcal{B})} \times \frac{1.64\times 10^{-13}}{a} \;(\textrm{TeV}\;\textrm{cm}^2\,\textrm{s})^{-1}
\label{equ:fluxpsr}
\end{equation}

where $\mathcal{B}$ from 1 to 9 is the analysis bin, $E_\mathcal{B}$ is the data excess in bin $\mathcal{B}$, $M_\mathcal{B}$ is the expected counts from simulations in bin $\mathcal{B}$, $a$ is the area of a HEALPix pixel, and $GHW_\mathcal{B}$ is the weight in bin $\mathcal{B}$ based on the signal-to-background ratio.
\end{enumerate}

After generating the all-sky flux density map, we then bin the pixels in Galactic longitude or latitude to obtain the profiles of of the Galactic diffuse emission.

\section{Contributions from Unresolved Sources}

In the study of diffuse emission, a major difficulty is to disentangle the true diffuse emission from unresolved sources. To estimate contributions from unresolved sources below our detection threshold, we use a method so-called number-intensity relation or log\,\textit{N}-log\,\textit{S} distribution \cite{casanova}. 

\subsection{Number-Intensity Relation for the 2HWC sources}

\begin{figure}[htpb]
\begin{center}
\includegraphics[width=0.55\linewidth]{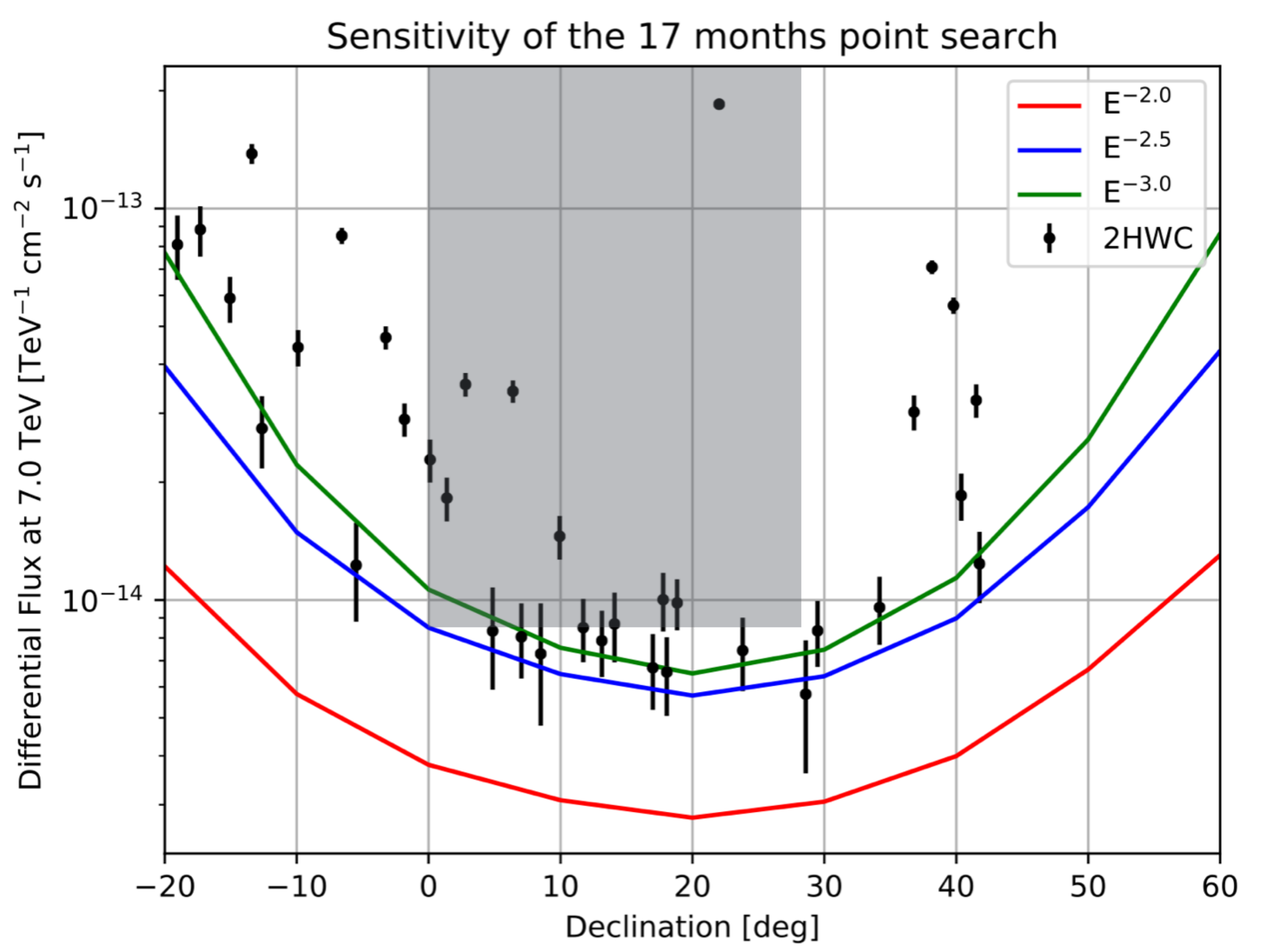}
\caption{Sensitivity at 7 TeV with 17 months of HAWC data for three spectral hypotheses, as a function of declination. Black points are the flux normalization and statistical uncertainty of 2HWC sources \cite{2hwc}. The sources in the shaded area are used in this analysis.}
\label{fig:sensi}
\end{center}
\end{figure}

Number-intensity relation, i.e. log\,\textit{N} (>\textit{S})-log\,\textit{S} where \textit{S} represents the source flux and \textit{N} the number of sources with their fluxes higher than \textit{S}, is a good method to examine the collective properties of a source population and, in our case, to estimate the contributions from unresolved sources that are below our detection threshold. However, the sensitivity of HAWC is a function of declination as shown in Figure \ref{fig:sensi}. In order to collect an unbiased population, only sources with a declination between $0^\circ$ and $28^\circ$ are selected, which approximately corresponds to a range of Galactic longitude between $33^\circ$ and $65^\circ$. In this region, the sensitivity of HAWC is better than $8.5\times 10^{-15}(\textrm{TeV}\,\textrm{cm}^2\,\textrm{s})^{-1}$ at 7 TeV and relatively flat, corresponding to $\sim5\%$ of the Crab flux \cite{crab}. This value is used as the threshold in our source selection. The region with Galactic longitude $>65^\circ$ is excluded because of the Cygnus region. Table \ref{table:flux} lists 14 2HWC sources in this region.

\begin{table}
  \centering
  \caption{Sources from the 2HWC catalog used in this analysis \cite{2hwc}.} \label{table:flux}
{\small
  \begin{tabular}{ccccccc}
    \hline
      RA & Dec & l & b & Flux at 7 TeV & Flux Uncertainty & \multirow{2}{*}{Spectral Index}\\
      (deg) & (deg) & (deg) & (deg) & $\times10^{-15}(\textrm{TeV}\,\textrm{cm}^2\,\textrm{s})^{-1}$ & $\times10^{-15}(\textrm{TeV}\,\textrm{cm}^2\,\textrm{s})^{-1}$ & \\
    \hline
282.39 & 0.11 & 32.82 & 0.47 & 22.8 & 2.9 & -2.54 \\
283.01 & 1.38 & 34.23 & 0.50 & 18.2 & 2.3 & -2.90 \\
284.33 & 2.80 & 36.09 & -0.03 & 35.5 & 2.5 & -2.93 \\
285.51 & 4.86 & 38.46 & -0.14 & 8.3 & 2.4 & -3.22 \\
286.79 & 8.50 & 42.28 & 0.41 & 7.3 & 2.5 & -3.25 \\
287.05 & 6.39 & 40.53 & -0.80 & 34.1 & 2.2 & -2.52 \\
288.11 & 9.93 & 44.15 & -0.08 & 14.5 & 1.9 & -2.93 \\
288.68 & 11.72 & 46.00 & 0.25 & 8.5 & 1.6 & -2.83 \\
290.30 & 13.13 & 47.99 & -0.50 & 7.9 & 1.5 & -2.75 \\
290.70 & 14.09 & 49.01 & -0.38 & 8.7 & 1.8 & -2.49 \\
292.15 & 17.78 & 52.92 & 0.14 & 10 & 1.7 & -2.56 \\
292.63 & 18.84 & 54.07 & 0.24 & 9.8 & 1.5 & -2.74 \\
294.74 & 23.81 & 59.37 & 0.94 & 7.4 & 1.6 & -2.96 \\
297.42 & 24.46 & 61.16 & -0.85 & 19.4 & 4.2 & -2.38 \\
\hline
  \end{tabular}
}
\end{table}

\begin{figure}[htpb]
\begin{center}
\includegraphics[width=0.6\linewidth]{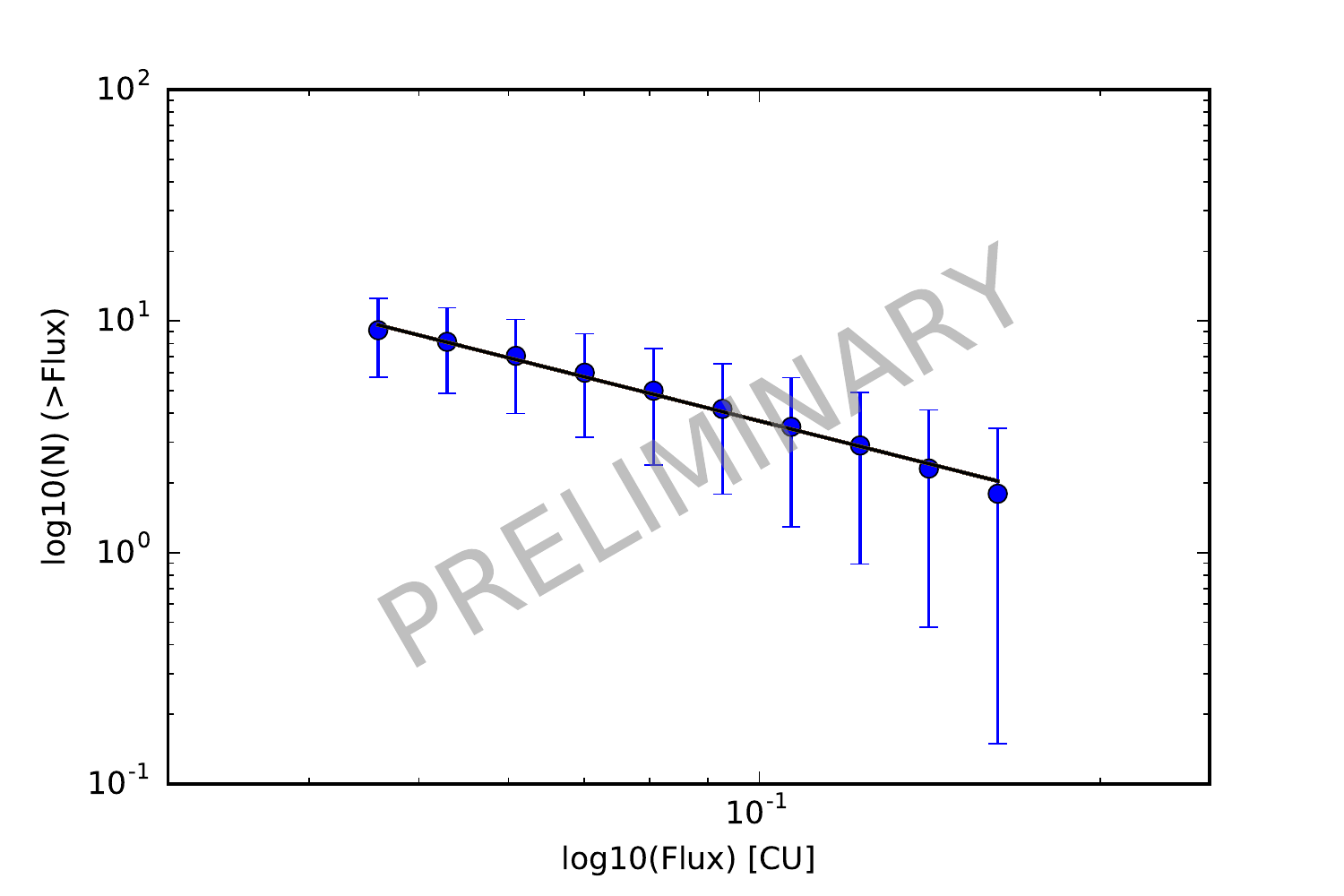}
\caption{Number-intensity relation from the 2HWC catalog in the region of Galactic longitude between $33^\circ$ and $65^\circ$. The statistical and systematical uncertainties on source fluxes are included using a Monte-Carlo based method. X-axis: Logarithmic differential flux at 7 TeV in Crab Unit. Y-axis: Logarithmic number of sources above a certain flux value. Black: best fit with a power law index of $-1.2\pm0.4$.}
\label{fig:logn-logs}
\end{center}
\end{figure}

Each source has its statistical uncertainty on the flux normalization, and additionally 50\% of systematic uncertainty. In order to take into account the uncertainties, a Monte-Carlo based procedure was used. For each source, a random flux is generated assuming a Gaussian error distribution with statistical and systematic uncertainties added in quadrature. The log\,\textit{N}-log\,\textit{S} distribution is then generated based on each set of random fluxes from all sources. The sources with flux less than $8.5\times 10^{-15} (\textrm{TeV}\,\textrm{cm}^2\,\textrm{s})^{-1}$ at 7 TeV are also included as their fluxes may fluctuate up to above our threshold given the statistical and systematical uncertainties. This procedure is repeated for a thousand times and the mean and standard deviations of \textit{N} are found. Finally, a Poisson error is added in quadrature to the standard deviation in each bin to account for the uncertainty due to the small number of sources. The log\,\textit{N}-log\,\textit{S} distribution from this analysis is shown in Figure \ref{fig:logn-logs}, where the black line is the best fit with a power law function. The index is $-1.2\pm0.4$, slightly softer but still consistent with -1, the expectation for uniformly distributed sources on a thin disk.

\subsection{Distribution of Pulsars and SNRs in the Galaxy}

In this section, we will calculate the expected number-intensity index based on the distribution of Galactic TeV sources in our Galaxy and compare the expectation with the result obtained in section 4.1 base on the 2HWC sources.

\begin{figure}[htpb]
\begin{center}
\includegraphics[width=0.45\linewidth]{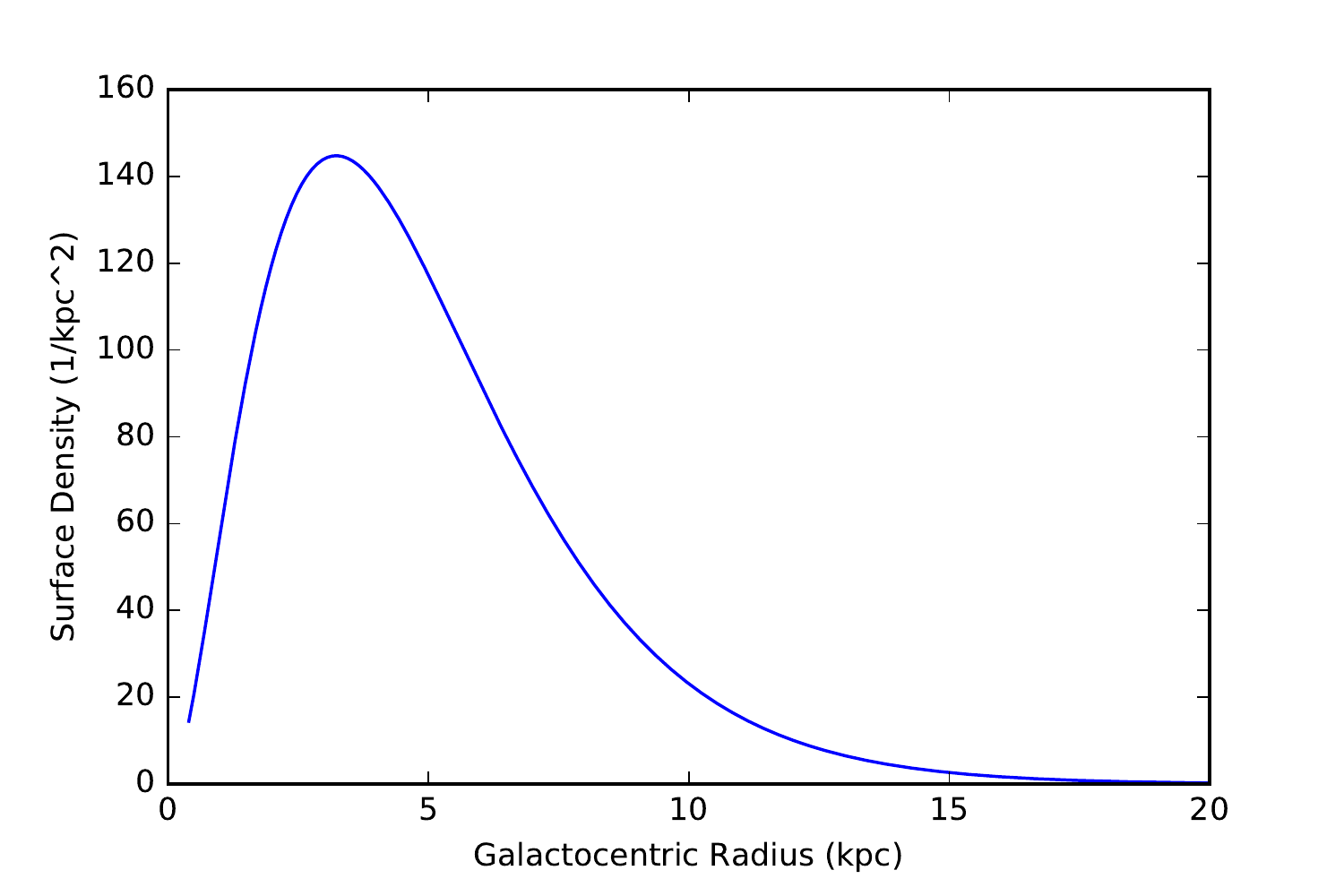}
\includegraphics[width=0.45\linewidth]{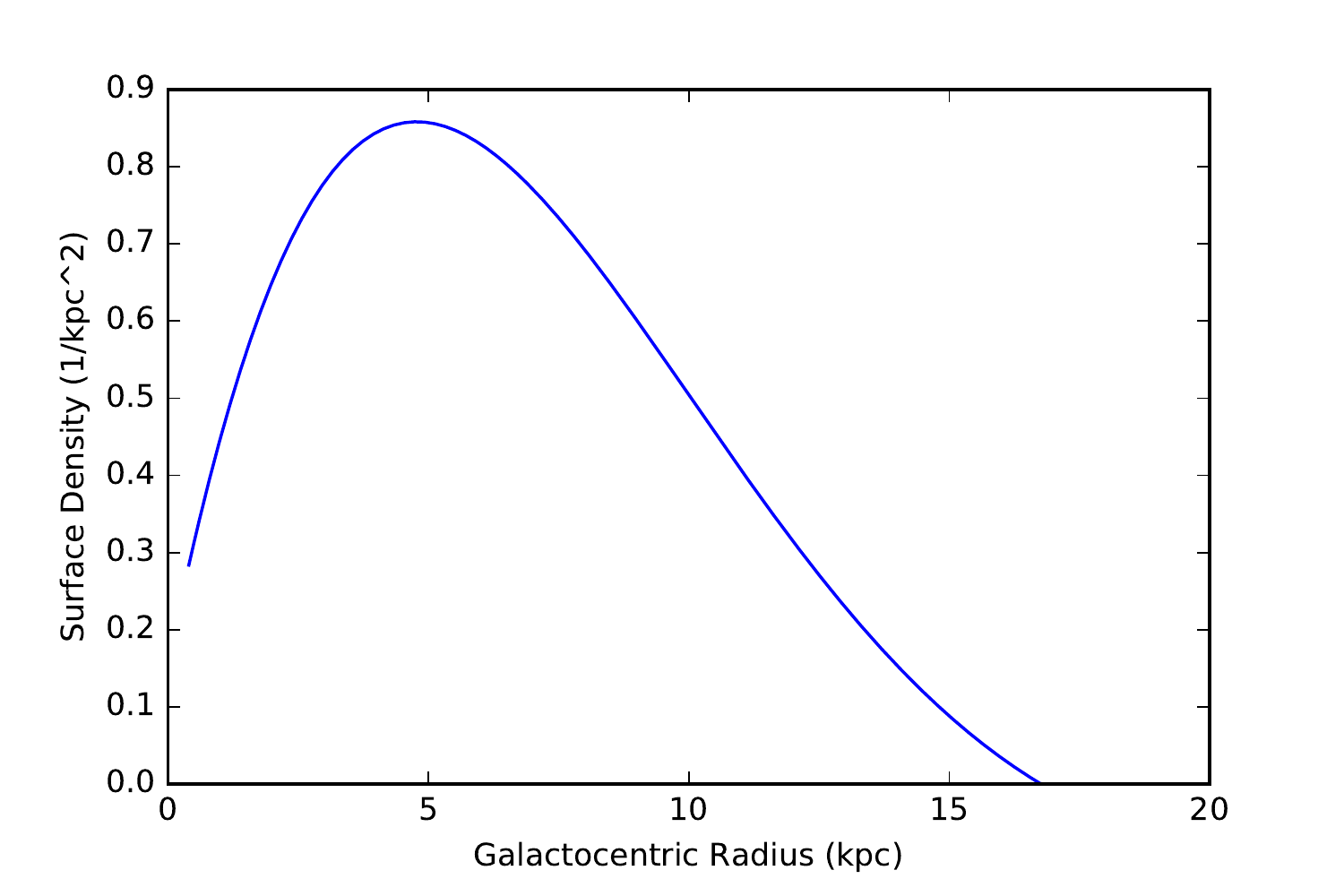}
\caption{Surface density of pulsars (left panel) and SNRs (right panel) as a function of Galactocentric radius.}
\label{fig:gcr}
\end{center}
\end{figure}

Pulsar wind nebulae and supernova remnants (SNRs) are two major types of Galactic TeV sources. The distributions of their counterparts in radio wavelengths, pulsars and SNRs, are well known. From observations at radio wavelengths, the surface density of pulsars $\sigma_{PSR}(r)$, where r is the Galactocentric radius, is 

\begin{equation}
\sigma_{PSR}(r) = a\Big(\frac{r}{r_0}\Big)^b e^{\big[-c(\frac{r-r_0}{r_0})\big]}
\label{equ:psrr}
\end{equation}
where $a = 41\,\textrm{kpc}^{-2}$, $b = 1.9$, and $c = 5.0$ \cite{lorimer}. The distribution along the height $z$ over the Galactic plane is

\begin{equation}
\sigma_{PSR}(z) = d e^{-\frac{|z|}{e}}
\label{equ:psrz}
\end{equation}
where $d = 0.75$ and $e = 0.18$ kpc. The surface density of SNRs $\sigma_{SNR}(r)$ follows

\begin{equation}
\sigma_{SNR}(r) = 
  \begin{cases}
    \sigma_{0SNR}(r) sin(\frac{\pi r}{r_2} + \theta_0) e^{-\beta r}, & \textrm{for}\;r<16.8 \,\textrm{kpc}\\
    0, & \textrm{for}\;r>16.8\,\textrm{kpc}
  \end{cases}
\label{equ:snrr}
\end{equation}
where $\sigma_{0SNR}(r)=(1.96\pm1.38) \,\textrm{kpc}^{-2}$, $r_2 = (17.2\pm1.9) \,\textrm{kpc}^{-2}$, $\theta_0 = (0.08\pm0.33)$, and $\beta = (0.13\pm0.08)$ \cite{green}. Along the $z$ axis, it follows the same exponential distribution as for the pulsars but with parameter $d = 0.58$ and $e = 0.083$ kpc \cite{xu}. Figure \ref{fig:gcr} shows the surface density of pulsars and SNRs as a function of Galactocentric radius. As both pulsars and SNRs have narrow distributions along the $z$ axis, in the following analysis, we assume these sources are distributed in a thin disk on the Galactic plane.

\begin{figure}[htpb]
\begin{center}
\includegraphics[width=0.45\linewidth]{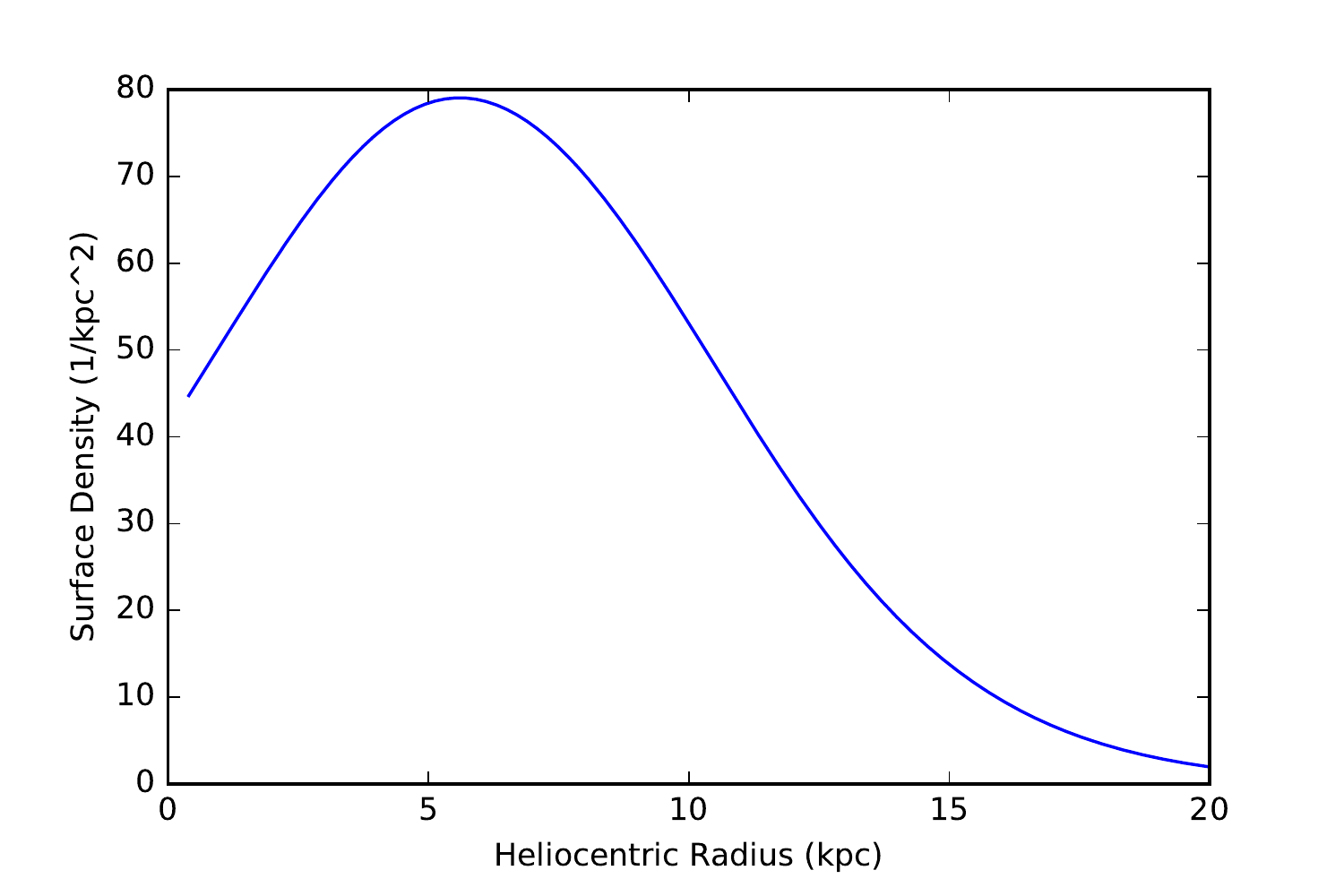}
\includegraphics[width=0.45\linewidth]{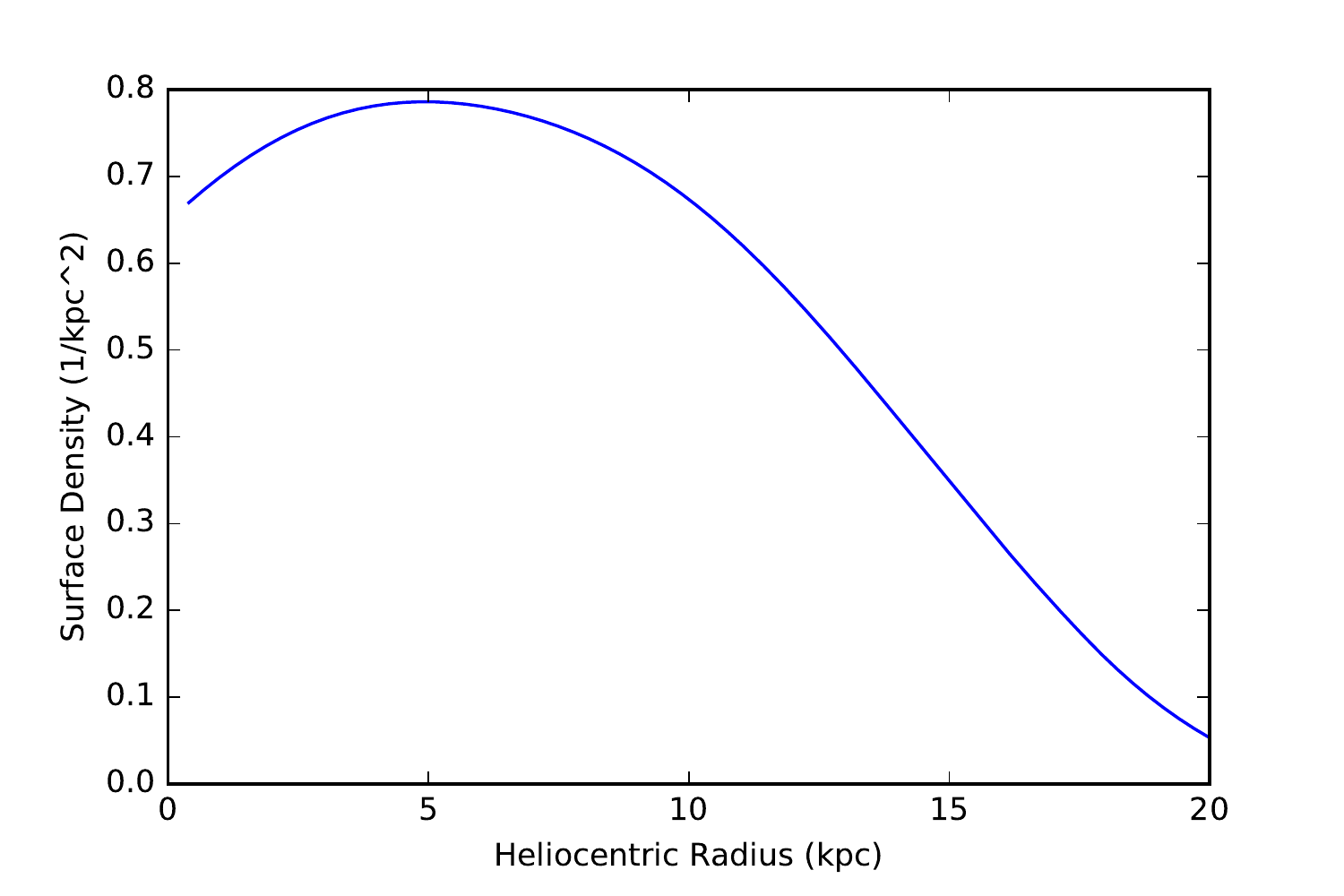}
\caption{Surface density of pulsars (left panel) and SNRs (right panel) as a function of heliocentric radius in the region of Galactic longitude between $33^\circ$ and $65^\circ$.}
\label{fig:hcr}
\end{center}
\end{figure}

Given the distance of $\sim8$ kpc from the sun to the Galactic center \cite{gc}, we can rewrite the surface density as a function of heliocentric radius. Figure \ref{fig:hcr} shows the surface density of pulsars and SNRs with Galactic longitude between $33^\circ$ and $65^\circ$, where the HAWC sensitivity is rather flat.

\begin{figure}[htpb]
\begin{center}
\includegraphics[width=0.55\linewidth]{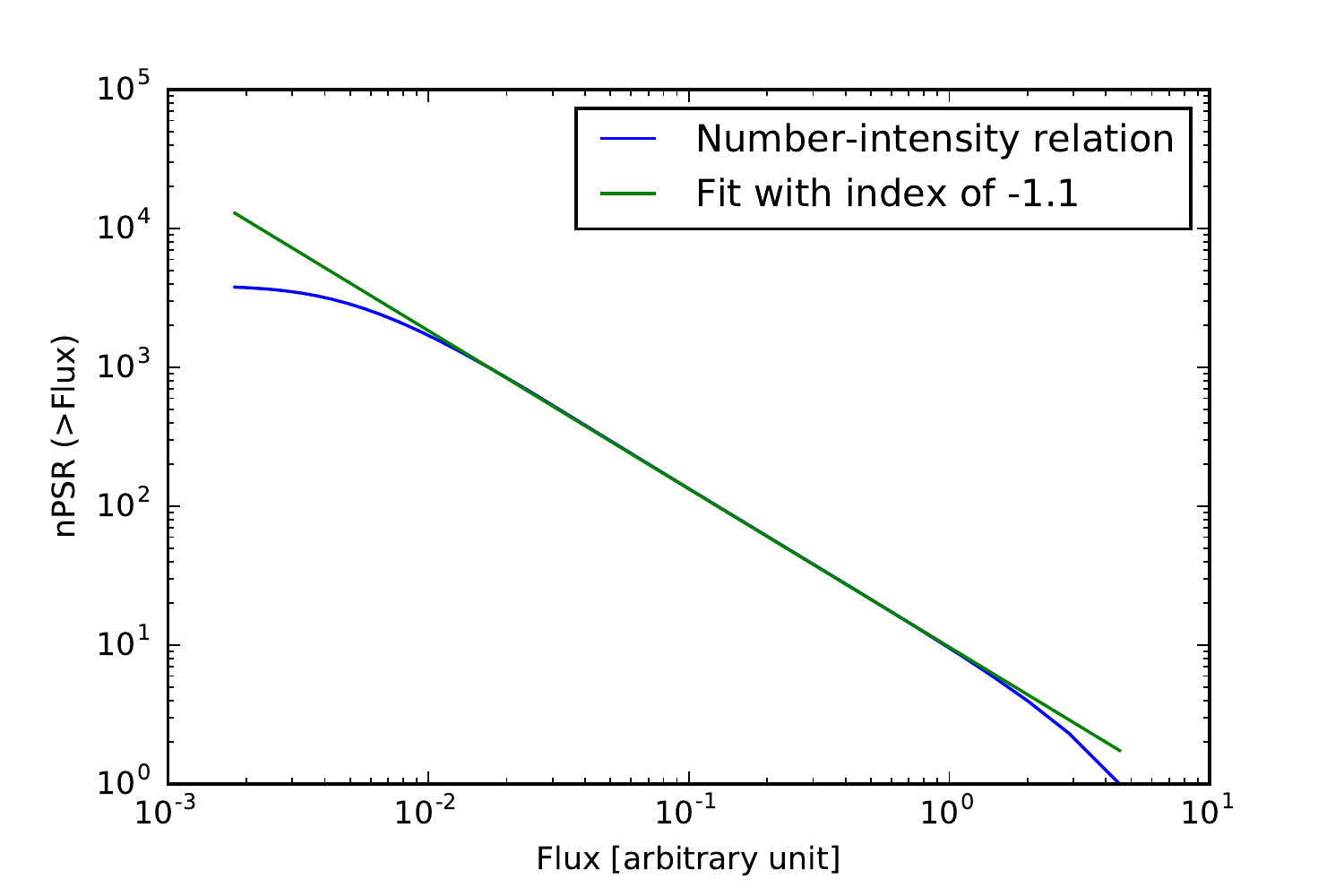}
\caption{Blue: Expected number-intensity relation from the pulsar population in the region of Galactic longitude between $33^\circ$ and $65^\circ$. Green: best fit with a power law index of -1.1.}
\label{fig:expni}
\end{center}
\end{figure}

Assuming all the pulsars have similar luminosity, the flux of a source is proportional to $1/\textrm{distance}^2$. The expected number-intensity relation from the pulsar population in the region of Galactic longitude between $33^\circ$ and $65^\circ$ is shown in Figure \ref{fig:expni}. The number-intensity relation is well fit with a power law with an index of -1.1. It is slightly softer than the index of -1 from uniformly distributed sources on a thin disk%, as there are more dim sources at large distances toward the Galactic center
. The SNR population yields a similar distribution. 

It turns out that considering the distribution of pulsars and SNRs does not alter the number-intensity index much compared to a uniform disk model. The index of $-1.2\pm0.4$ obtained from data is consistent with both expected indices from the pulsar and SNRs distributions in our Galaxy and from a uniform disk model. However, the error bar on the measured index is large. 

Including a larger fraction of the sky with more resolved sources will help better constraining the number-intensity index. However, the detector efficiency at different declinations must be corrected. The study of correcting the detect efficiency is ongoing and will provide a better constraint on the number-intensity index and on the flux contribution from unresolved sources.

%\section{Results}
%The results of this analysis will be presented at the ICRC2017.

\section*{Acknowledgments}
We acknowledge the support from: the US National Science Foundation (NSF); the
US Department of Energy Office of High-Energy Physics; the Laboratory Directed
Research and Development (LDRD) program of Los Alamos National Laboratory;
Consejo Nacional de Ciencia y Tecnolog\'{\i}a (CONACyT), M{\'e}xico (grants
271051, 232656, 260378, 179588, 239762, 254964, 271737, 258865, 243290,
132197), Laboratorio Nacional HAWC de rayos gamma; L'OREAL Fellowship for
Women in Science 2014; Red HAWC, M{\'e}xico; DGAPA-UNAM (grants IG100317,
IN111315, IN111716-3, IA102715, 109916, IA102917); VIEP-BUAP; PIFI 2012, 2013,
PROFOCIE 2014, 2015;the University of Wisconsin Alumni Research Foundation;
the Institute of Geophysics, Planetary Physics, and Signatures at Los Alamos
National Laboratory; Polish Science Centre grant DEC-2014/13/B/ST9/945;
Coordinaci{\'o}n de la Investigaci{\'o}n Cient\'{\i}fica de la Universidad
Michoacana. Thanks to Luciano D\'{\i}az and Eduardo Murrieta for technical support.

\end{document}